\documentstyle[prl,aps,multicol,psfig]{revtex}

\begin{document}

\draft

\title{Density of states and reflectionless tunneling in NS junction
  with a barrier}

\author{M. Schechter, Y. Imry and Y. Levinson}

\address{Department of Condensed Matter Physics, The Weizmann
  Institute of Science, Rehovot 76100, Israel}

\date{October 21, 1997}

\maketitle
   
\begin{abstract} 
The effect of the barrier on the proximity effect in normal-superconductor
junction is analyzed. A general criterion for the barrier, though large, to be
effectively transparent, is given. This criterion is applied to both the
conductance of a disordered NS junction (reflectionless tunneling) and the
density of states across it, showing that both phenomena stem from the same
physical effect.  
\end{abstract}
    
\begin{multicols}{2}

At the interface between a superconductor and a normal metal Cooper pairs in
the superconductor are being transformed to electrons and holes in the normal
metal by the process of Andreev reflection \cite{Andreev}.  This causes the
pair amplitude $\langle \psi_\downarrow \psi_\uparrow \rangle$ to be finite in
the normal metal close to the superconductor (the proximity effect).  The
existence of a barrier at the interface reduces the probability of Andreev
reflection, and therefore weakens the proximity effect.

The effect of the barrier on the conductance of an NS junction (reflectionless
tunneling) and on the density of states (DOS) across the junction was studied 
intensively (see e.g. 
\cite{Golubov89,van92,Volkov93,Volkov94,Hekking94,Beenakker94,Zhou95}).  
It was found that the condition
for the barrier to substantially reduce the conductance of the system is
$\Gamma \ll l_n/d$, where $\Gamma$ is the transmission probability through the
barrier, $l_n$ is the mean free path and $d$ is the length of the normal metal
between the interfaces with the superconductor and the reservoir.  van Wees
{\it et al.} \cite{van92} used a path integral picture to explain this effect. 
They show that when the disorder in the normal metal generates multiple
reflections from the interface, the constructive quantum interference which
results from the phase conjugation between electrons and holes enhances the
probability of Andreev reflection.  We follow this treatment and show that the
general criterion for the barrier not to be effective is that trajectories that
reflect from the interface more than $\Gamma^{-1}$ times before electron-hole
coherence is lost are dominant.  We then show how this criterion is applied to
the different cases of reflectionless tunneling and local 
DOS with and without magnetic field, to give the specific conditions for the
barrier to be effective in each.  We show that in all the different situations
this criterion reduces to the condition $\Gamma \ll l_n/\xi$, where $\xi$ is a
characteristic length which is energy, $\epsilon$, and magnetic field, $H$,
dependent, and defines the distance from the interface, at which electrons and
holes are still coherent.  For the problem of the DOS this gives the same
condition obtained by Volkov \cite{Volkov94} and later by Zhou {\it et al.}
\cite{Zhou95}.  This criterion can be applicable to other systems as well,
including special cases of ballistic systems as the one considered by Hekking
and Nazarov \cite{Hekking94}.

In the next section we give a simple and complete treatment of the effect of
the barrier on the DOS in an NS junction, using quasiclassical Green function
formalism. Then we use path integral picture and obtain analytically the
conductance in the limit of zero temperature, voltage and magnetic field. 
Finally, we use the path integral picture to explain all the above phenomena on
the same grounds.

{\it Density of States in NS structure with a barrier.} The model we consider
is an NS slab of width $W \leq \xi_s$ in the $y$ direction ($\xi_s =
\sqrt{\hbar D_s/\Delta}$ is the superconducting coherence length, $D_s$ is the
superconductor diffusion constant, $\Delta$ is the superconductor order
parameter).  The normal metal and the superconductor are semi infinite in the
$x$ direction (perpendicular to the interface).  We restrict ourselves to the
dirty limit, $\hbar/\Delta > \tau_n,\tau_s$, where $\tau_n$ and $\tau_s$ are
the normal state elastic mean free times of the normal metal and the
superconductor.

Since we are considering the dirty limit we use the Usadel equation
\cite{Usadel}.  We neglect inelastic scattering and scattering from
paramagnetic impurities and use the simplest model without self-consistency
taking $\Delta$ to be constant in the superconductor and zero in the normal
metal.  We consider small magnetic field, $W \ll l_B$ ($l_B$ is the magnetic
length), in which case the derivative of $\Theta$ in the $y$ direction can be
neglected, and after integrating the equation in the $y$ direction we obtain a
one dimensional equation for $\Theta(\epsilon,x)$ \cite{Bruder96}
\begin{equation} 
\frac{d^2}{dx^2} \Theta = \frac{-2 i \epsilon}{D} \sin \Theta
- \frac{2 \Delta} {D} \cos \Theta + \frac{1}{2} \gamma \sin 2 \Theta ,
\label{eq:Usadel} \end{equation} 
where $\Theta(\epsilon,x)$ is defined by the
retarded Green functions: $G_{\epsilon} = \cos\Theta, F_{\epsilon} =
\sin\Theta$, and $\gamma = e^2 H^2 W^2/3 = W^2/3 l_B^4$.  After solving the
equation, the local DOS is obtained from $N(\epsilon,x) = N_0
ReG_{\epsilon}(x)$, $N_0$ is the normal metal DOS at the Fermi energy.

We consider first the case $H = 0$.  Applying the boundary conditions
at $+\infty$ (normal metal) and $-\infty$ (superconductor) one obtains
a general solution to Eq.\ (\ref{eq:Usadel}) \cite{Bruder96}
\begin{equation} 
  \Theta(\epsilon,x) =  \left\{ \begin{array}{ll} 
      4 \tan^{-1}\{\tan(\frac{\psi_n}{4}) 
      \exp[-x/\zeta_n(\epsilon)]\} , 
      &  x > 0 \\ 
      \tilde{\Theta} + 4 \tan^{-1}\{\tan(\frac{\psi_s}{4}) 
      \exp[x/\zeta_s(\epsilon)]\} , 
      & x < 0 . 
    \end{array}   
  \right.
  \label{eq:generalNS}
\end{equation} 
Here $\zeta_n = \sqrt{i/2} \xi_n$, where $\xi_n = \sqrt{\hbar
  D_n/\epsilon}$ is the coherence length of electron-hole pairs in the
normal metal, which determines the energy dependent length scale of
the proximity effect in the normal metal, and diverges at small
energies ($D_n$ is the diffusion constant in the normal metal).
$\zeta_s = \xi_s (1 -\epsilon^2/\Delta^2)^{-1/4}$ is the decay length
for states with $\epsilon < \Delta$ in the superconductor.  The
constants $\psi_n$ and $\psi_s$ are to be determined by the two
boundary conditions at the NS interface, and $\tilde{\Theta} =
\tan(\Delta/-i \epsilon)$ is the value of $\Theta$ in a bulk
superconductor.

One of the boundary conditions for the Usadel equation
\cite{Zaitsev84,Kuprianov88} is a consequence of current conservation,
and is therefore independent of the existence and strength of the
barrier at the interface.  This boundary condition reads: $\sigma_s
d/dx(\Theta(\epsilon, 0_-)) = \sigma_n d/dx(\Theta(\epsilon, 0_+))$,
where $\sigma_n$ and $\sigma_s$ are the normal state conductances of
the N and S metals.  The second boundary condition depends on the
transmission probability of the barrier, and in the limit of small
transmission reads \cite{Kuprianov88}:
\begin{equation} 
  l_n \frac{d}{dx} \Theta(\epsilon, 0_+) = \tilde{\Gamma} 
  \sin(\Theta(\epsilon, 0_+) - \Theta(\epsilon, 0_-)) . 
  \label{eq:boundary2} 
\end{equation} 
Here $\tilde{\Gamma}$ is a measure of the transparency of the barrier,
and is given by $\tilde{\Gamma} = \frac{3}{2} \left \langle
  \cos\varphi \Gamma_{\varphi} /(1 - \Gamma_{\varphi}) \right
\rangle$, where $\Gamma_{\varphi}$ is the angular dependent
transmission probability of the barrier, $\varphi$ is the angle of
incidence at the barrier, and the averaging is over the total solid
angle.  Using the above boundary conditions and defining: $\eta =
\sigma_n \xi_s / \sigma_s \xi_n , \,\,\, \beta = l_n/\zeta_n , \,\,\,
\alpha = \left (-i \epsilon/\sqrt{\Delta^2 - \epsilon^2} \right
)^{1/2}$, we obtain two equations for $\psi_s$ and $\psi_n$:
\begin{equation} 
  -2 \beta \sin(\psi_n/2) = \tilde{\Gamma} \sin(\psi_n - \psi_s - 
  \tilde{\Theta}) , 
  \label{eq:firstbound}
\end{equation}
\begin{equation} 
  \sin(\psi_s/2) = - \eta \alpha \sin(\psi_n/2) . 
  \label{eq:secondbound}
\end{equation} 
From Eq.\ (\ref{eq:firstbound}) and the definitions of $\beta$ and
$\zeta_n$ we see that the important parameter of the problem is the
ratio between $\tilde{\Gamma}$ and the parameter $l_n / \xi_n$
measuring the "dirtiness" of the normal metal \cite{Zhou95}.

\begin{figure} 
	\narrowtext 
	\centerline{\psfig{figure=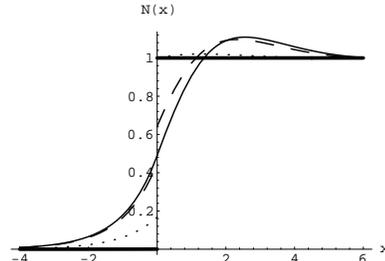,width=2in}}
	\caption{Density of states as a function of distance from 
	the NS boundary in units of $\xi_s$ (negative coordinate values 
	correspond to the superconductor).$\epsilon = 0.2 \Delta$ in all 
	curves. Different curves correspond to different ratios of 
	$\tilde{\Gamma}/|\beta|$. Thick line - 0, dotted - 0.4,  dashed - 
	5, thin line - $\infty$ (no barrier). $\eta = 1$ in all curves. }
	\label{fig-dosall}
\end{figure}

In the limit of infinite barrier ($\tilde{\Gamma}/|\beta| = 0$),
$\psi_n = 0$, and using Eq.\ (\ref{eq:secondbound}) $\psi_s = 0$, and
the DOS on each side of the interface equals its bulk value (zero for
the superconductor, $N_0$ for the normal metal, see Fig.\ 
\ref{fig-dosall}, thick line).  When the transmission of the barrier
is small compared to the "dirtiness" ($\tilde{\Gamma}/|\beta| \ll 1
$), $\psi_n$ and $\psi_s$ are small, we obtain a large jump in the DOS
across the barrier, and the deviation from the infinite barrier result
is of first order in $\tilde{\Gamma}/|\beta|$ in the superconducting
side, and of second order in $\tilde{\Gamma}/|\beta|$ in the normal
metal side (Fig.\ \ref{fig-dosall}, dotted line).  In the limit of
$|\beta|/\tilde{\Gamma} = 0$ Eq.\ (\ref{eq:firstbound}) takes the form
of $\psi_n = \psi_s + \tilde{\Theta}$, which is the boundary condition
for the case of no barrier at the boundary, and since the second
boundary condition does not depend on the barrier, the results for the
local DOS in this case are equivalent to those obtained for the case
of no barrier \cite{Bruder96,Gueron96}, even though $\Gamma \ll 1$ (Fig.\ 
\ref{fig-dosall}, thin continuous line).  For $|\beta|/\tilde{\Gamma}
\ll 1$ we have $\psi_n - \psi_s - \tilde{\Theta} \ll 1$, and the local
DOS is close to the no barrier limit (Fig.\ \ref{fig-dosall}, dashed
line).  This means that in this limit the barrier, though large, has
only a small effect on the DOS in the system.  Since $\xi_n$ is energy
dependent and diverges at low energies, there is always an energy
region where the limit $\tilde{\Gamma} > l_n/\xi_n$ is satisfied and
the barrier is not effective. In the case $\tilde{\Gamma} > l_n/\xi_s$
the barrier is not effective for all energies smaller than $\Delta$. 

In the next sections we show that the barrier is not effective in
certain limits due to the coherence of electrons and holes in the
normal metal. Therefore, it is expected that an applied magnetic field
will destroy the effect.  We first consider the effect of finite
magnetic field at the Fermi energy.  In the normal metal the Usadel
equation can be solved exactly, and the solution is $\Theta = 2
\tan^{-1}[\tan(\psi_n/2)\exp(-\sqrt{\gamma} x)]$.  In order to solve
the Usadel equation analytically in the superconductor we consider
small magnetic fields, $H \ll \Phi_0/W \xi_s$ {\it i.e.} $\xi_H \gg
\xi_s$, where $\Phi_0$ is the flux quantum and $\xi_H$ is defined by
$H W \xi_H = \Phi_0$.  We then use the boundary condition
(\ref{eq:boundary2}), and consider $\tilde{\Gamma} \ll l_n/\xi_s$.  We
see that the barrier is not effective for $\tilde{\Gamma} \gg
l_n/\xi_H$, and for $\tilde{\Gamma} \ll l_n/\xi_H$ the effect is
destroyed (the barrier becomes effective) \cite{Volkov94}.

With both finite magnetic field and finite energy the Usadel equation
is not solved exactly in the normal metal. However, multiplying the
equation by $d \Theta /dx$, integrating over $x$ and using the
boundary condition at infinity, we obtain the first order equation:
\begin{equation} 
  \frac{d \Theta}{dx} = - [\frac{-4 i \epsilon}{D} (1 - \cos\Theta) + 
  \frac{1}{2} \gamma ( 1 - \cos2\Theta)]^{\frac{1}{2}} . 
\end{equation} 
Using the boundary condition (\ref{eq:boundary2}) we obtain an
equation for the values of $\Theta$ at the interface on both sides:
\begin{eqnarray}
  l_n[\frac{-4 i \epsilon}{D} (1 - \cos\Theta_n(0)) + 
  \frac{1}{2} \gamma ( 1 - \cos2\Theta_n(0))]^{\frac{1}{2}} \nonumber \\ 
  = \tilde{\Gamma} \sin(\Theta_s(0) - \Theta_n(0)) .
\end{eqnarray} 
\noindent 
To obtain analytical results we assume $\epsilon \ll \Delta$ and $H
\ll \Phi_0/W \xi_s$, and then it can be shown that $\Theta_s(0)
\approx \pi/2$.  The condition for the barrier to be noneffective is:
$\tilde{\Gamma} \gg l_n (\xi_n^{-4} + \xi_H^{-4})^{1/4}$.  This
condition reduces to the conditions found above for the case of no
magnetic field and the case of finite magnetic field at zero energy.
We find that in magnetic field and energy space there is a line given
by $(\epsilon/\epsilon_c)^2 + (H/H_c)^4 = 1$, ($\epsilon_c = D_n
\tilde{\Gamma}^2 / l_n^2$, $H_c = \tilde{\Gamma} \Phi_0/l_n W$) that
separates the domain where the barrier is not effective (small
$\epsilon$ and $H$) from the domain where the energy and magnetic
field destroy the effect, and the barrier is effective.

\begin{figure} 
	\narrowtext 
	\centerline{\psfig{figure=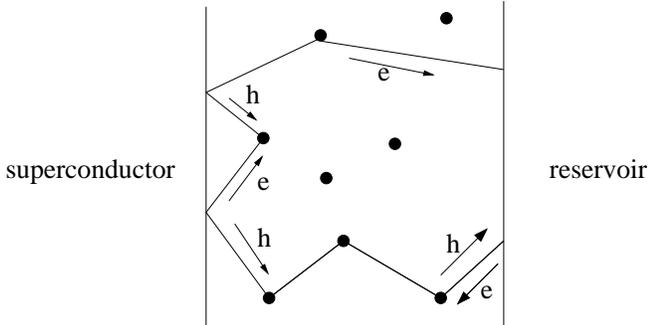,width=3.375in}}
	\caption{Geometry of the model, an example of a trajectory with 
	$N$=2}
	\label{fig-van}
\end{figure}

{\it Reflectionless tunneling.} The phenomenon of reflectionless tunneling,
related to the experimental observation of excess low voltage conductance in NS
junctions (see, e.g.  \cite{Kastalsky91,Nguyen92}), is treated in many
theoretical works (e.g.
\cite{Golubov89,van92,Volkov93,Volkov94,Hekking94,Beenakker94,Zhou95}).  A
physically transparent approach to reflectionless tunneling was given by van
Wees {\it et al.} \cite{van92}.  They discuss the phenomenon of reflectionless
tunneling in a semiclassical model, in which the motion of the electrons and
holes in the normal metal follows a deterministic trajectory which depends on
the initial position and direction of the electron, but the barrier between the
normal metal and the superconductor is treated quantum mechanically.  The
validity of the semiclassical treatment is discussed both in their paper and by
Beenakker and van Houten \cite{Houten91}.

van Wees {\it et al.} consider an N'NS system with a barrier at the NS
interface. N' is an electron reservoir.  Assuming, at first, a totally
reflecting barrier at the interface with the superconductor, an
electron leaving the reservoir into the disordered normal metal at a
given position and direction has a given trajectory in the normal
metal till it will reach the reservoir again.  This trajectory can not
hit the barrier at all or hit the barrier any number of times $N$ (see
Fig.\ \ref{fig-van}). 

For a disordered normal metal of length $d$ the transmission
probability of electrons across it is $T=l_n/d$, and therefore, the
fraction of trajectories which return to the reservoir after hitting
the barrier $N$ times is:
\begin{equation}
  F(N) = \left\{ \begin{array}{ll} 
      1-T  &  N=0 , \\ 
      T^2 (1-T)^{N-1}  &  N \neq 0 . 
    \end{array}   
\right.
\label{eq:FN} 
\end{equation}
Considering a barrier with finite transmission probability $\Gamma$,
at each reflection from the interface either normal or Andreev
\cite{Andreev} reflections are possible.
 
When $\epsilon=0, H=0$, the Andreev reflected hole retraces exactly
the path of the incoming electron. Therefore, an incoming electron can
return to the reservoir either as an electron at the end of the
trajectory, or as a hole at the beginning of the trajectory (Fig.\ 
\ref{fig-van}).

We define $r_{ee}(N)$ and $r_{hh}(N))$ as the amplitudes of electron
and hole reflections in an $N$ trajectory, respectively.  The average
contribution to the charge current of trajectories involving $N$
reflections at the interface is $I(N) = 2 |r_{he}(N)|^2$ and therefore
the conductance of the system is given by: $G(V \rightarrow 0,H = 0) =
G_s \sum_{N=0}^{\infty} F(N) I(N)$, where $G_s$ is the Sharvin
conductance of the interface, and for a system with a finite
cross-section $G_s = 2 e^2 n/\hbar$, $n$ is the number of transverse
channels.  Considering first the case where $\epsilon = 0$, $H = 0$,
we obtain a recurrence formula: $r_{he}(N)=i(|r_{he}(N-1)| +
|r_{he}|)/(1 + |r_{he}| |r_{he}(N-1)|)$, where $r_{he}$ is the
amplitude for the Andreev reflection of an electron at the interface,
given by $r_{he} = r_{eh} = i \Gamma/(2 - \Gamma)$.  The above formula
is obtained by summing the amplitudes of all the different paths
within the trajectory that result in a hole being reflected back to
the reservoir (a path is a given sequence of normal and Andreev
reflections from the different points the trajectory intersects the
interface. There are infinitely many paths for a given trajectory).
 
We then obtain: $r_{he}(N) = i \,\, \tanh(N \tanh^{-1}(|r_{he}(1)|))$
and for a barrier with small transmission probability
\begin{equation}
  r_{he}(N) \approx i \,\, \tanh(N |r_{he}(1)|) \approx i \,\, 
  \tanh(N \Gamma/2 ) . 
  \label{eq:rheapprox}
\end{equation} 
Inserting this result in the conductance formula we obtain:
\begin{equation}
  G(V \rightarrow 0,H = 0) = \left\{ \begin{array}{ll} 
      \frac {2 e^2 n}{h} \frac{\Gamma ^2}{T}  &  (\Gamma \ll T) , \\ 
      \frac {2 e^2 n}{h} (\frac{1}{2T} + \frac{1}{\Gamma})^{-1}  &  (\Gamma 
      \gg T) . 
    \end{array}   
  \right. 
  \label{eq:Glimits}
\end{equation}
This result differs from the result obtained by Beenakker {\it et al.}
\cite{Beenakker94} by a factor of 2 in the $\Gamma \gg T$ limit since
they consider the case of short range disorder while we consider the
case where the electrons and holes follow a classical trajectory,
which prevails when the scattering potential varies slowly on the
scale of a wavelength \cite{Houten91}.

The strong dependence of the amplitude of Andreev reflection on the
number of times the trajectory hits the interface, as manifested in
Eq.\ (\ref{eq:rheapprox}), is due to the coherence of the incoming
electron and reflected hole. Due to this coherence, the probability of
Andreev reflection, $|r_{he}(N)|$ approaches 1 for $N \gg
\Gamma^{-1}$, and the probability of normal reflection, $|r_{ee}(N)|$
approaches zero. The phenomenon of reflectionless tunneling is a
result of the fact that as the disorder is enhanced, the fraction of
large $N$ trajectories grows.

{\it Reflectionless tunneling and density of states.} Both
reflectionless tunneling and the influence of the barrier on the DOS
in an NS structure are related to the coherent transport of pairs from
the superconductor across the barrier to the normal metal, which is
given by the probability of an electron approaching the boundary to be
Andreev reflected from it.  The probability of Andreev reflection
approaches 1 for trajectories with $N \gg \Gamma^{-1}$, so the barrier
is not effective if large $N$ trajectories occur with high
probability.  Therefore, the general criterion for the barrier not to
be effective is that trajectories that hit the barrier more than
$\Gamma^{-1}$ times before losing the coherence between holes and
electrons are dominant in the system.  Using random walk theory it can
be shown that the length of a trajectory between $N$ consecutive times
it hits the barrier is of the order of $\bar{L}_N \approx N^2 l_n$
\cite{Feller}, and therefore only when coherent trajectories with
lengths larger than $L_{\Gamma} \equiv l_n/\Gamma^2$ occur with high
probability will the barrier not be effective.  This criterion can be
applied to various cases, where different mechanisms limit the length
of coherent trajectories.

In the case of reflectionless tunneling the length of the normal
metal, $d$, is what limits the trajectories to lengths of order
$d^2/l_n$ and therefore the barrier is not effective when $\Gamma \gg
T$.  In the case of the DOS in NS structure with semi-infinite normal
metal, there is no limit to the length of the trajectories, and
therefore, when $\epsilon=0$ and $H=0$, the barrier is not effective.
However, at finite $\epsilon$ and $H$ the electron-hole coherence is
limited.  The equation for $r_{he}(2)$ is generalized \cite{van92}:
$r_{he}(2) = [r_{he} (1 + \exp(i \Delta \phi)]/[1 + |r_{he}| |r_{eh}|
\exp(i \Delta \phi)]$.  Here $\Delta \phi = 2 \epsilon \bar{L}/\hbar
v_F + 4 \pi (H A/\Phi_0)$, the first part is due to the phase
accumulated by an electron and a hole traversing a part of a
trajectory of length $\bar{L}$ between the two intersection points
with the barrier, and the second part is due to the magnetic flux, $H
A$, through the Andreev loop consisting of the part of the trajectory
of length $\bar{L}$, and closed by the superconductor.  For electrons
at $\epsilon=0,H=0$, $\Delta \phi = 0$, $r_{he}(2) = 2 r_{he}/(1 +
|r_{he}|^2)$ and the probability $|r_{he}(2)|^2$ has a maximum.  In
the same way $|r_{he}(N)|^2$ (\ref{eq:rheapprox}) has a maximum when
$\Delta \phi = 0$.  When electron-hole coherence breaks ($\Delta \phi
\approx 2\pi$) there is no enhancement in the Andreev reflection
probability due to constructive interference, and the barrier becomes
effective.  At finite energies coherence will be maintained up to
lengths of order $L_c = \hbar v_F/\epsilon$, and therefore the DOS
only at $\epsilon \ll \Gamma^2 \hbar/\tau_n$ will not be affected by
the barrier ($L_{\Gamma} \ll L_c$).  This is the same condition
($\Gamma \gg l_n/\xi_n$) which was obtained using the Usadel equation.

At $\epsilon = 0, H \neq 0$, it can be shown that in a slab of width
$W$ a trajectory of length $l_B^4/W l_n$ encloses one flux quantum and
therefore the condition for the barrier not to be effective is $\Gamma
\gg l_n/\xi_H$, again as was obtained using the Usadel equation.
 
Similar considerations lead to the result that reflectionless
tunneling is destroyed at energies of the order of the Thouless
energy, $\hbar D_n/d^2$, and magnetic fields of the order of $H =
\Phi_0/Wd$.

The criterion of hitting the barrier more than $\Gamma^{-1}$ times
before losing the electron-hole coherence can also be applied to
ballistic systems where the geometry allows returns to the interface,
as an SININ double barrier system \cite{Alban96}, or the system
considered by Hekking and Nazarov \cite{Hekking94}.

We benefited from stimulating discussions with W. Belzig.  We also
thank C. Bruder, A. Krichevsky, Z. Ovadyahu and F. Wilhelm for useful
discussions.  This work was supported by the Israel Academy of Science
and by the German-Israeli Foundation (GIF).

\end{multicols}
\end {document}